\documentclass[floatfix,showpacs,amsmath,amssymb,letterpaper,groupaddresses,superscriptaddress]{article}
\usepackage{latexsym}
\usepackage{epsfig}

\usepackage{amsmath}
\usepackage{amssymb}
\usepackage{graphicx}
\usepackage{times}
\usepackage[section]{placeins}
\usepackage{caption}
\usepackage{subcaption}
\usepackage{cite}

\usepackage{color}

\def\ar{\arrowvert}

\def\be{\begin{equation}}
\def\ee{\end{equation}}
\def\ba{\begin{eqnarray}}
\def\ea{\end{eqnarray}}
\def\bc{\begin{center}}
\def\ec{\end{center}}

\def\nn{\nonumber}

\def\h{\hskip 1cm}
\def\hm{\hskip 0.5cm}
\def\hmm{\hskip 0.2cm}
\def\hM{\hskip 0.1cm}

\def\ra{\rangle}
\def\la{\langle}

\begin{document}
	
	\vspace{4cm}
	
	\begin{center}
		
		{\Large \bf Exact generation of quantum states by the dynamics of spin chains}\\
		
		\vspace{2cm}
		
		Morteza Moradi, \h Vahid Karimipour\\
		
		\vspace{1cm}
		
		Department of Physics, Sharif University of Technology, P.O. Box 11155-9161, Tehran, Iran.\\
		
	\end{center}
	
	\vspace{1cm}
	
	\begin{abstract}
We design a quasi-one-dimensional spin chain  with engineered coupling strengths such that the natural dynamics of the spin chain  evolves a single excitation localized at the left-hand  site to any specified single particle state on the whole chain. Our treatment is an exact solution to a problem which has already been addressed in approximate ways.  As two important examples, we study the $W$ states and Gaussian states with arbitrary width.

	\end{abstract}
	PACS numbers: 89.70.+c, 03.65
	
	\vskip 1cm

	\section{Introduction}\label{Intro}
	Quantum spin chains, apart from being an indispensable tool for understanding a large variety of phenomena in condensed matter physics, have also been a large laboratory for the investigation of exactly solvable models in many-body quantum systems. One of the main goals  in these disciplines has been to find specific quantum spin chains for which the ground state and correlation functions can be found in closed form.  With the upsurge of quantum computation and information theory, it is now almost a decade that the dynamics of spin chains has attracted   attention in connection with quantum information processing tasks \cite{Bose1,Bose2,ChDEL,BuB,Kay1,FPK1,FPK2,BaB,Kay2,BeB,KE,BACVV,RossKay,SAK,AK, Kay}. Starting with \cite{Bose1}, spin chains turned out to be excellent carriers of quantum states in short distances either with very high  or with perfect fidelity \cite{ChDEL,BuB,Kay1,FPK1,FPK2,BaB,Kay2}. Since then the plethora of quantum information tasks for quantum spin chains has considerably expanded, including entanglement distribution \cite{Jur, lat, BACVV}, measurement-based quantum computation \cite{Rus, BeB,KE}, perfect routings \cite{RossKay,SAK, AK} and quite recently state generation \cite{Kay} which is the subject of the present paper. The importance of this problem, that is the capability of initializing a quantum register to any given state cannot be overemphasized. Such a problem will have many applications, i.e. in quantum simulations among other domains. Here the goal is to design a specific Hamiltonian which can evolve a single excitation which is completely localized on one site to a given desired state which is distributed over all spins. \\
	
	\noindent More precisely, given a state 
	\be\label{psi}
	|\psi\ra=\sum_{k=1}^N \psi_k|k\ra,
	\ee
	the idea is to design a Hamiltonian such that after a time $t_0$, we  have 
	\be
	|\psi\ra=e^{-iHt_0}|1\ra.\nn
	\ee
	Here $|k\ra$ means the state $|0,\cdots, 0, 1,0 
	\cdots 0
	\ra$ where only the spin on site $k$ is excited. The states $\{|k\ra, k=1\cdots N\}$  span the one-excitation sector of the Hilbert space. Naturally here we have in mind those Hamiltonians which conserve the number of excitations and hence commute with the total spin operator, i.e. $[H,S_z]=0.$
	A prototype of these Hamiltonians is the $XY$ Hamiltonian given by
	\be
	H= \sum_{n=1}^N\frac{B_n}{2}(1-
	Z_n)+\sum_{n=1}^{N-1} J_n (X_{n}X_{n+1}+Y_nY_{n+1}),\nn
	\ee
	where $X_n,Y_n,$ and $Z_n$ are the Pauli matrices acting on site $n$. Recently this problem was posed 
	and investigated in  \cite{Kay} , where it was shown that provided that no two consecutive amplitudes of $|\psi\ra$ are zero, the local magnetic fields $B_n$ and the local couplings $J_n$ can be engineered in such a way that $|\psi\ra$ can be generated with arbitrary precision. However, the actual values of  couplings $B_n$ and $J_n$  had to be found numerically and by iteratively tuning the Hamiltonian. As admitted in \cite{Kay} the disadvantage of this numerical method was that the time $t_0$ scaled as $N^2$, making the process rather slow. To remedy this, the author of \cite{Kay}  proposed an alternative analytical method which could produce only a very limited number of states. One could then hope that by using various perturbative techniques one could deform these states so that the given state can be approximated.  \\
	
		\noindent Our goal in this paper is to solve the problem of state generation analytically for all one-particle states in an exact way. To this aim, we utilize a quasi-one-dimensional chain shown in Fig.\ref{SpinNetFig}(a). The crucial point for this kind of geometry is that the chain decomposes into a direct sum of virtual chains of two spins for which the evolution of an excitation is extremely simple.  It is this decomposition and the subsequent simplicity of the dynamics which allows an exact determination of the couplings for all kinds of states. While in \cite{Kay}, this problem is connected to an inverse eigenvalue problem which is solved iteratively, here we solve the problem by exactly and simultaneously following the evolution of the particles (more precisely the probability amplitudes of a single particle) on all the small chains. This leads to a set of coupled non-linear equations for the couplings which we solve in closed form.   We should remind that there are other quasi-one-dimensional chains \cite{RossKay} which have a simple apparent geometry than the one shown in Fig.\ref{SpinNetFig}(a). However, they decompose into virtual chains of length two and three and it is not easy to simultaneously follow  the dynamics of the particles as described above and solve the subsequent non-linear equations.\\

	\noindent	In summary for any state of the form (\ref{psi}), we exactly determine the coupling constants $J_n$ and the times $t_n$ for applying the single qubit $Z_n$ gates. As examples, we consider the generation of $W$ states and  Gaussian states of various width on chains of different lengths. The results for these examples are shown in figures \ref{WSFig} and \ref{GSFig}. \\
	
		{\bf Remark:} We should emphasize that compared to the method of \cite{Kay}, which uses a time-independent Hamiltonian and generates a limited class of single-particle states, the price that we pay for this exact generation of all single-particle states,  is that we need to apply local single qubit $Z$ gates at specific times. This substitutes the tuning of local albeit static magnetic  fields on all sites proposed in \cite{Kay}. The extent to which the timing of these pulses is crucial for the success of the scheme is discussed in section (\ref{error}). \\

		\noindent The structure of the paper is as follows: in section \ref{SpinNetStr} we simply analyze the structure of the quasi-one-dimensional chain and its equivalence to the virtual 2-chains and examine the dynamics of the chain. In section \ref{EvoSpinNet} we determine the coupling constants and the times for applying the Z-pulses. Section \ref{Ex} is devoted to examples where we study two important classes of examples,  namely the W-states and the Gaussian states. We end the paper with an outlook.

	\section{The spin network structure} \label{SpinNetStr}
	
\begin{figure}[h]
		\centering
		\includegraphics[width=1\textwidth]{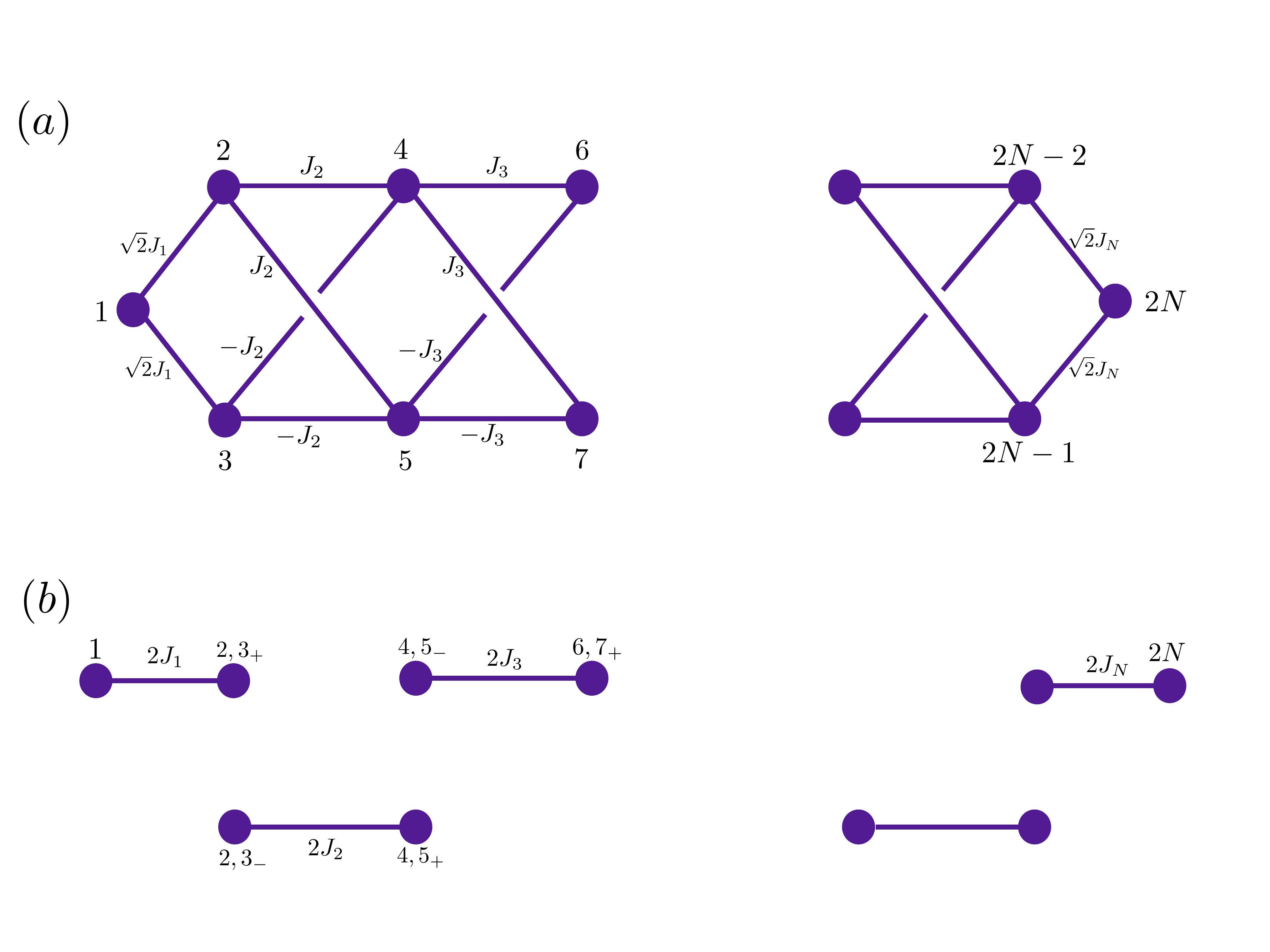}
		\caption{
			(a) A spin network containing two 1-D chains with regular interaction between them. The vertices represent qubits, and the edges used to show XX coupling between qubits of strength $J_k$.
			(b) N virtual spin chains of length 2 equivalent to the spin network in (a). The coupling strengths are shown on the edges. In our examples, we use the lower part of the network, the sites with odd index, only as ancillary qubits. That is the amplitude on all these sites are zero and the state is supported only on the above leg of the chain, i.e. on the even-numbered spins. }
		\label{SpinNetFig}
	\end{figure}

	We introduce the  spin network shown in Fig.\ref{SpinNetFig}(a) where each link entails a Hamiltonian $h:=\frac{1}{2}(X\otimes X + Y\otimes Y)$ with strength $J$ written on the link. As is seen in each block all the interaction of horizontal and oblique links are equal modulo the signs. It is known that in architectures based on Josephson Junction superconducting qubits, which are modeled by XX Hamiltonians, it is possible to implement couplings with negative signs \cite{jos}. The main point is that in the one-particle sector the Hamiltonian $h$ is nothing but a simple hopping term. In fact, it is well known and easily verified that 
\be
h_{i,j}=\frac{1}{2}(X_iX_j+Y_iY_j)=|i\rangle\langle j|+|j\rangle\langle i|.
\ee
	Therefore $h|0,0\ra=h|1,1\ra=0$, $h|0,1\ra=|1,0\ra$ and $h|1,0\ra=|0,1\ra.$
	
\noindent	As the XX Hamiltonian commutes with $Z_n$:
	\begin{center}
		$[H,\sum_{i=1}^{N}{Z_i}]=0$,
	\end{center}
	 if we start from a single excitation in site 1 or any other site, the dynamics will be confined in the one-particle sector. One can now consider an arbitrary block like the one containing the spins $2,3,4$ and $5$. The part of Hamiltonian pertaining to these spins can be rewritten as 
	
		\ba\label{2345}
			H_1&=&J_2{({\ar2\ra\la4\ar}+{\ar2\ra\la5\ar}-{\ar3\ra\la4\ar}-{\ar3\ra\la5\ar})+h.c.}\cr &=&J_2{(\ar2\ra-\ar3\ra)(\la4\ar+\la5\ar)+h.c.}		=2J_2(\ar2,3_-\ra\la4,5_+\ar)+h.c.,
    \ea
	where $\forall i,j\in\{1,2,...,2N\}:\ar i,j_{\pm}\ra:=\frac{1}{\sqrt{2}}(\ar i\ra \pm \ar j\ra)$. The same analysis applies to the next block whose Hamiltonian is rewritten as 
	\ba\label{4567}
		H_2&=&J_3{({\ar4\ra\la6\ar}+{\ar4\ra\la7\ar}-{\ar5\ra\la6\ar}-{\ar5\ra\la7\ar})+h.c.}\cr&=&J_3{(\ar4\ra-\ar5\ra)(\la6\ar+\la7\ar)+h.c.}=
		2J_3(\ar4,5_-\ra\la6,7_+\ar)+h.c..
	\ea
	Noting that all the states written in the right-hand side of (\ref{2345}) and (\ref{4567}) are orthogonal to each other, it turns out that the chain decomposes into a collection of spin chains of length 2 shown in Fig.\ref{SpinNetFig}(b). With the definitions  $\ar0,1_-\ra:=\ar 1\ra,\ar 2N,2N+1_+\ra:=\ar2N\ra$, the final Hamiltonian can be written as a collection of independent 2-spin chains, as in Fig.\ref{SpinNetFig}(b):
	\be
	H=\sum_{n=1}^{N}{2J_n{\ar2n-2,2n-1_{-}\ra\la2n,2n+1_+\ar}}+h.c.
	\ee

	\vspace{1cm}
	
	\section{Dynamics in the spin network} \label{EvoSpinNet}
	If we were to use this chain for perfect state transfer, our task would be more straightforward. We only needed to move a single excitation from site 1 to site $23_+$ and then apply a Z-pulse to site 3 to move the excitation  from the site $23_+$ to $23_-$ and put it on the beginning of the next chain which automatically goes over to the end of this site after a certain time and then repeat the process until the excitation reaches the other end of the total chain. However, in  generating states we want to distribute the excitation with prescribed probabilities all over the chain and hence also all over the virtual chains. This is a much harder task than state transfer in which when the excitation leaves a virtual 2-spin chain, we do not need to take it into account  anymore. Here as times passes we have to know how  all the excitations in all the 2-chains (more precisely  the probabilities of a single excitation in all the 2-chains) evolve in time. This is where the dynamics of a 2-spin chain, compared with a 3-spin chain plays a crucial role. Denoting the two sites of a 2-chain simply by 1 and 2, we have 
	\bc
	$H=J\boldsymbol{X_1}.\boldsymbol{X_2}\equiv \frac{J}{2}(X_1\otimes X_2 + Y_1\otimes Y_2)=J(\ar1\ra\la2\ar+\ar2\ra\la1\ar)=
	\begin{bmatrix}
	0&J\\
	J&0
	\end{bmatrix}$,
	\ec
	and hence  
	\be
	e^{-iHt}\ar1\ra=cos(Jt)\ar1\ra-i\hM sin(Jt)\ar2\ra.
	\ee

	\noindent Let us start from the state $|1\ra$. The dynamics of the chain evolves this state after time $t_1$ within the leftmost chain:
	
	\be
	e^{-iHt_1}\ar1\ra=cos(2J_1t_1)\ar1\ra-i\hM sin(2J_1t_1)\ar2,3_+\ra\nn
	\ee
	Applying the $Z_3$ gate at time $t_1$ turns this into 
	\be
	Z_3e^{-iHt_1}\ar1\ra=cos(2J_1t_1)\ar1\ra-i\hM sin(2J_1t_1)\ar2,3_-\ra.\nn
	\ee
	The excitation is now on both  site 1 of the first 2-chain and site $23_-$ of the second 2-chain with the indicated amplitudes.  After time $t_2$ both amplitudes evolve and  after the pulse $Z_5$ we have 
	\ba
	Z_5e^{-iHt_2}Z_3e^{-iHt_1}\ar1\ra&=&cos(2J_1t_1)\bigg[cos(2J_1t_2)\ar1\ra-i\hM sin(2J_1t_2)\ar2,3_+\ra\bigg]\cr
	&&-i\hM sin(2J_1t_1)\bigg[cos(2J_2t_2)\ar2,3_-\ra-i\hM sin(2J_2t_2)\ar4,5_-\ra\bigg].\nn
	\ea
	We can continue in this manner to find the state of the chain under the following dynamics
	
	\be
	|\psi\ra=	e^{-iHt}\ar1\ra:=e^{-iHt_N}Z_{2N-1}e^{-iHt_{N-1}}...Z_5e^{-iHt_2}Z_3e^{-iHt_1}\ar1\ra.
	\ee
	To find the amplitudes in a simpler way, a descriptive way is very effective: 
		 After the pulse $Z_3$ which is applied at $t_1$, a fraction $-i\sin(2J_1t_1)$ is at the beginning of the second chain, namely on $23_-$. After the pulse $Z_5$ which is applied at $t_2$, a fraction $-i\sin(2J_2t_2)$ of this amplitude moves to the beginning of the third chain, namely on site $4,5_-$. Continuing in this way, after the pulse $Z_{2k-1}$ which is applied at time $t_{k-1}$, the excitation has reached the site $(2k-2,2k-1)_-$ with an amplitude $$(-i\sin(2J_1t_1))(-i\sin(2J_2t_2))\cdots (-i\sin(2J_{k-1}t_{k-1}))=:(-i)^{k-1}A_{k-1}.$$ With the next pulse at site 
	$Z_{2k+1}$ a fraction $-i\sin(2J_kt_k)$ of this amplitude leaves this chain and a fraction $\cos (2J_kt_k)$ remains in the chain. It is now important that all the other applied pulses on sites $2k+3, 2k+5, \cdots$ do not affect this amplitude which hereafter changes only by the internal dynamics of the short chain $[(2k-2,2k-1)_-, (2k,2k+1)_+]$. Thus the explicit form of the wave function is given by:

	\ba\label{EvoEq}
    |\psi\ra&=&\sum_{k=1}^{N}(-i)^{k-1}A_{k-1}cos(2J_k(\tau_{k}-\tau_{k+1}))cos(2J_k\tau_{k+1})\ar2k-2,2k-1_-\ra\nn \\
	&&+\sum_{k=1}^{N}(-i)^{k}A_{k-1}cos(2J_k(\tau_{k}-\tau_{k+1}))sin(2J_k\tau_{k+1})\ar2k,2k+1_+\ra.
	\ea
	where 
	\be\label{Tau}
	\tau_k:=\sum_{i=k}^{N}t_k\hm;\hm\tau_{N+1}:=t_N,
	\ee
	and
	\be\label{A}
	A_k:=\prod_{i=1}^{k}sin(2J_it_i)\hm;\hm A_0:=1.
	\ee
	
\noindent	Now, we have to calculate the times and coupling strengths such that our intended arbitrary state (\ref{psi}) will be generated. For the time being, we focus on the absolute squares of all the coefficients in (\ref{psi}) as positive.  Once the state with the required probabilities is generated on the  chain, we can apply  phase gates $e^{iZ_k\phi_k}$ to  tune also the local phases. 
	
	\subsection{Calculating the times and coupling strengths}
	Consider now a  state $|\psi\ra$ with some given amplitudes on the virtual chains. In this section, we first calculate the times $t_k$ and coupling strengths $J_k$ to create this state. We will then relate both these amplitudes and the corresponding times and coupling strengths to the actual quasi-one-dimensional chain in Fig.\ref{SpinNetFig}(a). 
	
	\subsubsection{Given probability amplitudes  in the virtual spin chains}
\begin{figure}[h]
		\centering
		\includegraphics[width=1\textwidth]{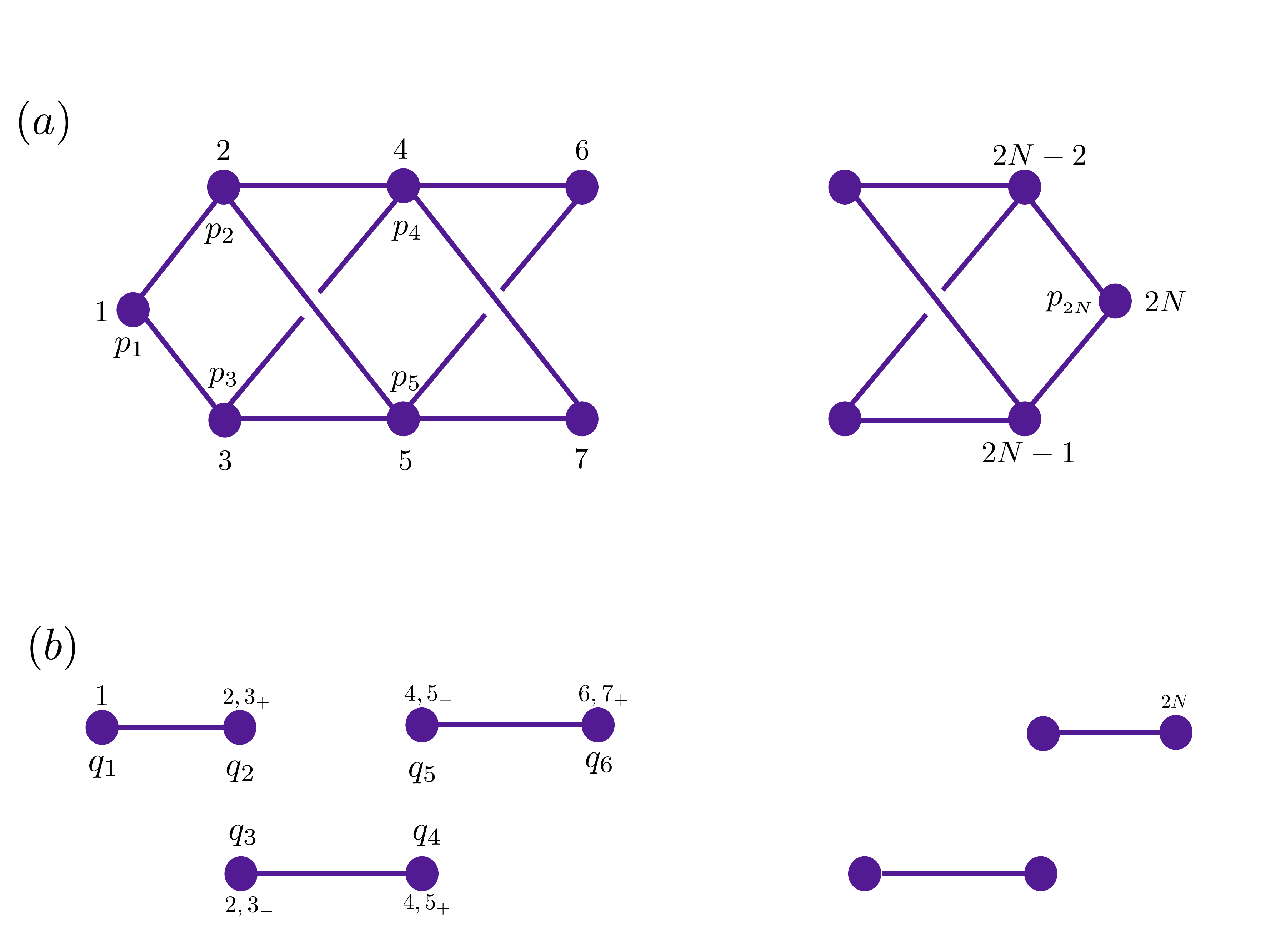}
		\caption{
			The site probabilities on the actual chain are denoted by $p_k$, while those on the virtual chains by $q_k$.  The probabilities on the virtual sites are determined from the probabilities on the actual sites above them, i.e.   $q_2$ and $q_3$ are  determined by $p_2$ and $p_3$ and so on as in (\ref{prob1}) and (\ref{qEq}).}
		\label{SpinNetworksProb}
	\end{figure}

	We first consider the virtual chain. The  probabilities  on the sites of this chain are denoted by $\{q_k\}$ and those on the actual chain are denoted by $\{p_k\}$, figure (\ref{SpinNetworksProb}). Suppose that $q_{2k+1}$ and $q_{2k}$ are respectively the probabilities  that   $\ar2k,2k+1_-\ra$ and $\ar2k,2k+1_+\ra$ are excited. Thus from the equation (\ref{EvoEq}):
	
	\ba\label{ProbEq}
	q_{2k-1}&=&[A_{k-1}cos(2J_k(\tau_{k}-\tau_{k+1}))cos(2J_k\tau_{k+1})]^2\\
	q_{2k}&=&[A_{k-1}cos(2J_k(\tau_{k}-\tau_{k+1}))sin(2J_k\tau_{k+1})]^2,\nn
	\ea
	where $A_k$ is given in (\ref{A}). From this set of coupled nonlinear equations, we should determine all the times $t_k$ and all the coupling constants $J_k$. 
	First, we divide the second equation of (\ref{ProbEq}) by the first to obtain 
	$
	\tan^2(2J_k\tau_{k+1})=\frac{q_{2k}}{q_{2k-1}},$
	or
	\be\label{new}
	\cos^2(2J_k\tau_{k+1})=\frac{q_{2k-1}}{q_{2k-1}+q_{2k}}.
	\ee
	{\bf Remark:} In case that two consecutive probabilities $q_{2k-1}
$ and $q_{2k}$ are zero, we only need to set $2J_k(\tau_k-\tau_{k+1})=\frac{\pi}{2}+m\pi$ and choose the parameter $2J_k\tau_{k+1}=N\pi$ see the explanation before equation (\ref{JTau1}).   Therefore in contrast to the method in \cite{Kay}, such states can also be generated by our method. \\

\noindent	Second, the sum of the two equations in (\ref{ProbEq}) leads to 
		\be\label{SumProbEq}
	q_{2k-1}+q_{2k}=A_{k-1}^2cos^2(2J_k(\tau_{k}-\tau_{k+1})).
	\ee
	Using (\ref{SumProbEq}) we find,
	
	\be
	\begin{aligned}
		A^2_k\hm &=&\prod_{i=1}^{k}sin^2(2J_i(\tau_{i}-\tau_{i+1}))=A^2_{k-1}sin^2(2J_k(\tau_{k}-\tau_{k+1}))\\
		&=&A^2_{k-1}(1-cos^2(2J_k(\tau_{k}-\tau_{k+1})))=A^2_{k-1}-(q_{2k-1}+q_{2k}).\nn
	\end{aligned}
	\ee
	By repeating this argument and using $$A^2_1=1-q_1-q_2$$ we find
	\be
	A^2_k=1-\sum_{i=1}^{2k}q_i=\sum_{i=2k+1}^{2N}q_i.\nn
	\ee
	
	\noindent This already leads to a very simple result:  despite its appearance as indicated in (\ref{A}), $A_k$ is a time-independent quantity which is solely determined by the probabilities. 
	From (\ref{SumProbEq}), we obtain  
	
	\be\label{JtEq0}
	\cos^2(2J_k(\tau_{k}-\tau_{k+1}))=\frac{q_{2k-1}+q_{2k}}{A_{k-1}^2}.
	\ee
	Equations (\ref{new}) and (\ref{JtEq0}), give the sequence of  ratios $\frac{\tau_{k+1}}{\tau_k}$ which finally leads to the determination of all $\tau_k$'s in terms of $\tau_1$ and then to the determination of all the coupling constants $J_k$. There are some important details, due to the multiple solutions of these equations,  which we describe below. \\
	
	\noindent Equation (\ref{ProbEq})  gives 
	
	\be\label{JTauEq}
	2J_k\tau_{k+1}=n_{k}\pi + cos^{-1}\sqrt{\frac{q_{2k-1}}{q_{2k-1}+q_{2k}}}\ ,\h\hmm  n_k\in \mathbb{Z}
    \h \ \ \ k=1,\cdots N,
	\ee
	where  integers $n_k$ are arbitrary. Also from (\ref{SumProbEq}) one finds 
		\be\label{JtEq}
	2J_k(\tau_{k}-\tau_{k+1})=m_k\pi+cos^{-1}\sqrt{\frac{q_{2k-1}+q_{2k}}{A_{k-1}^2}}\ ,\hmm\   m_k\in \mathbb{Z},\h k=1,\cdots N,
	\ee
	where again, the integers $m_k$ have to be chosen judiciously. 
	We will later argue that it is best to set the integers $m_k=0$ and $n_k=N$  to keep the couplings $J_k$ bounded.   Summing (\ref{JtEq}) and (\ref{JTauEq}) and setting $k=1$, we find 
	\be\label{JTau1}
	2J_1\tau_1=N\pi+cos^{-1}\sqrt{q_1+q_2}+cos^{-1}\sqrt{\frac{q_1}{q_1+q_2}}.
	\ee
	Naturally, this single equation does not yield the values of $J_1$ and $\tau_1$ independently since after all the time scale of the full dynamics can be tuned by the strength of the first coupling constant. However, from the two equations, all the other times and coupling constants can be determined.	
    Dividing (\ref{JTauEq}) by (\ref{JtEq}) and rearranging one finds:

	\be\label{Dividing}
	\frac{\tau_{k+1}}{\tau_k}=\frac{N\pi+cos^{-1}\sqrt{\frac{q_{2k-1}}{q_{2k-1}+q_{2k}}}}{N\pi+cos^{-1}\sqrt{\frac{q_{2k-1}+q_{2k}}{A_{k-1}^2}}+cos^{-1}\sqrt{\frac{q_{2k-1}}{q_{2k-1}+q_{2k}}}}.\nn
	\ee
	which after repeating and using (\ref{JTau1}) yields 
		\be\label{Tau_k+1}
	\tau_{k+1}=\frac{1}{2J_1}\frac{\prod_{i=1}^{k}[N\pi+cos^{-1}\sqrt{\frac{q_{2i-1}}{q_{2i-1}+q_{2i}}}]}{\prod_{i=1}^{k-1}[N\pi+cos^{-1}\sqrt{\frac{q_{2i+1}+q_{2i+2}}{A_{i}^2}}+cos^{-1}\sqrt{\frac{q_{2i+1}}{q_{2i+1}+q_{2i+2}}}]}.
	\ee
	
	\noindent From (\ref{JTauEq}) we can now determine all the coupling strengths:
	\be\label{JEq}
	J_k=J_1\prod_{i=1}^{k-1}\bigg[\frac{N\pi+cos^{-1}\sqrt{\frac{q_{2i+1}+q_{2i+2}}{A_{i}^2}}+cos^{-1}\sqrt{\frac{q_{2i+1}}{q_{2i+1}+q_{2i+2}}}}{N\pi+cos^{-1}\sqrt{\frac{q_{2i-1}}{q_{2i-1}+q_{2i}}}}\bigg].
	\ee

	\noindent We now determine the order of magnitude of couplings $J_k$. Since $\cos^{-1}(\dot)\in [0,\ \pi/2]$ according to Eq. (\ref{JEq}) for large N:
	
	\begin{equation}\label{BB1}
	\frac{J_k}{J_1}\le\prod_{i=1}^{k-1}[\frac{N\pi+\frac{\pi}{2}+
		\frac{\pi}{2}}{N\pi}]=(1+\frac{1}{N})^{k-1}<(1+\frac{1}{N})^N\simeq e,
	\end{equation}
	
	and
	
	\begin{equation}\label{BB2}
	\frac{J_k}{J_1}\geq\prod_{i=1}^{k-1}[\frac{N\pi}{N\pi+\frac{\pi}{2}}]=\frac{1}{(1+\frac{1}{2N})^{k-1}}>\frac{1}{(1+\frac{1}{2N})^{N}}\simeq \frac{1}{\sqrt{e}}.
	\end{equation}
	In deriving these bounds, the choice $m_k=0,\ n_k=\pi$ has played a crucial role and the result is that the order  of magnitude for $J_k$ and $J_1$ are  the same. Thus there is no exponential increase in the value of coupling constants or exponential decrease in the time interval between the pulses.  
	
	\subsubsection{Given probabilities in the spin network}
	The evolution of the spin network could be obtained from the evolution of the virtual spin chains. ‌By  inserting Eq. (\ref{ProbEq}) in Eq. (\ref{EvoEq}) we have:
	\be
	 e^{-iHt}\ar1\ra=\sum_{k=1}^{N}[(-i)^{k-1}\sqrt{q_{2k-1}}\ar2k-2,2k-1_-\ra+(-i)^k \sqrt{q_{2k}}\ar2k,2k+1_+\ra]\h\nn\\
	\ee
	 Using the definitions of  $\ar i,j_{\pm}\ra :=\frac{1}{\sqrt{2}}(\ar i\ra \pm \ar j\ra)$ this state is equivalent to the following state on the actual chain
	\be
	e^{-iHt}\ar1\ra=\sum_{k=1}^{N}\frac{(-i)^{k}}{\sqrt{2}}(\sqrt{q_{2k}}+\sqrt{q_{2k+1}})\ar2k\ra+\sum_{k=0}^{N-1}\frac{(-i)^{k}}{\sqrt{2}}(\sqrt{q_{2k}}-\sqrt{q_{2k+1}})\ar2k+1\ra.
	\ee
	Since we want to generate a state $\ar\psi_T\ra=\sum_{n=1}^{2N}\alpha_n\ar n\ra$ where $\ar\alpha_n\ar^2=P_n$, this gives 
	\be\label{prob1}
	P_{2k}=\frac{1}{2}(\sqrt{q_{2k}}+\sqrt{q_{2k+1}})^2\hm,\hm	P_{2k+1}=\frac{1}{2}(\sqrt{q_{2k}}-\sqrt{q_{2k+1}})^2
	\ee
	or equivalently:
	
	\be\label{qEq}
	q_{2k}=\frac{P_{2k}+P_{2k+1}}{2}+\sqrt{P_{2k}P_{2k+1}}\hm;\hm q_{2k+1}=\frac{P_{2k}+P_{2k+1}}{2}-\sqrt{P_{2k}P_{2k+1}}
	\ee
	Therefore for any set of given probabilities on the actual chain, one can immediately determine the probabilities on the virtual chain and then from (\ref{JEq}) and (\ref{Tau_k+1}) tune the coupling strengths and the pulse times to generate that give state. A minor simplification arises if we demand that the state has support only on the lower or upper branch of the quasi-one-dimensional chain, i.e. on the chain consisting of even-numbered qubits or odd-numbered qubits. In these cases where we use one of the branches as the main chain and the other branch as the ancilla chain, we are in fact dealing with a one-dimensional chain and in these cases, we have $P_{2k}P_{2k+1}=0$ and from (\ref{qEq}), we have $q_{2k}=q_{2k+1}$.

    \section{Examples}\label{Ex}
     In this section use the above mechanism to generate some well-known states. 
	 (1) W-states with equal probability of having an excited spin in each site and (2) Gaussian-states of different widths.\\
	 
	 {\bf Remark:} In our examples, we use the lower part of the network, the sites with odd index, only as ancillary qubits. That is the amplitude on all these sites are zero and the state is supported only on the above leg of the chain, i.e. on the even-numbered spins. So in both examples, the lower chain is empty and the state is generated on the above chain of even-numbered qubits.

	\subsection{W-states}
	
	For W-states the probabilities in the upper chain are equal and in the lower chain are zero:
	\be
	 \forall k\in\{1,2,...,N\}:\hmm P_{2k-1}=0 \hmm ,\hmm P_{2k}=\frac{1}{N}.\nn 
	\ee
	Therefore, we can find the probabilities in the virtual chains:
	\be
	q_1=0\hmm,\hmm\forall k\in\{1,2,...,N-1\}:\hmm q_{2k}=q_{2k+1}=\frac{1}{2N}\hmm,\hmm q_{2N}=\frac{1}{N}.	\nn
	\ee
	\begin{figure}[h]
	 	\centering
	 	\includegraphics[width=1\textwidth]{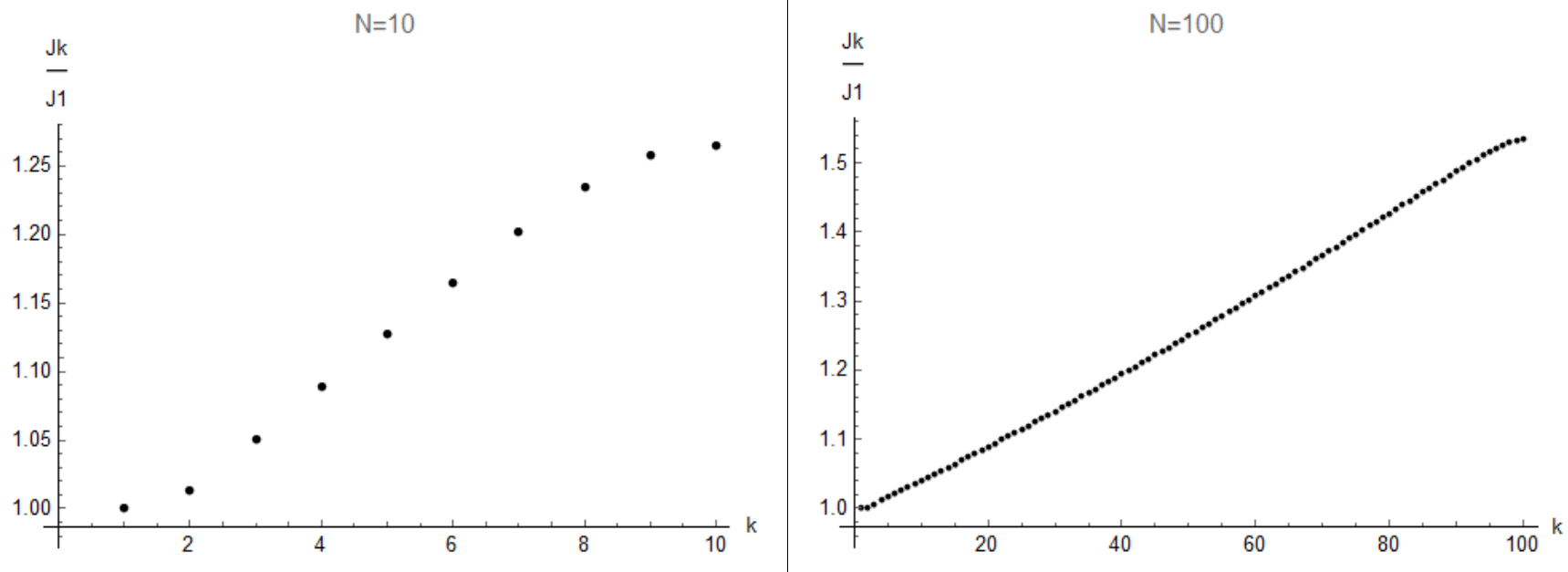}
	 	\caption{The coupling strengths for generating the W-states on chains of length   N=10 and N=100.}
	 	\label{WSFig}
	\end{figure}
 
 	\begin{figure}[h]
 \centering
 \includegraphics[width=1\textwidth]{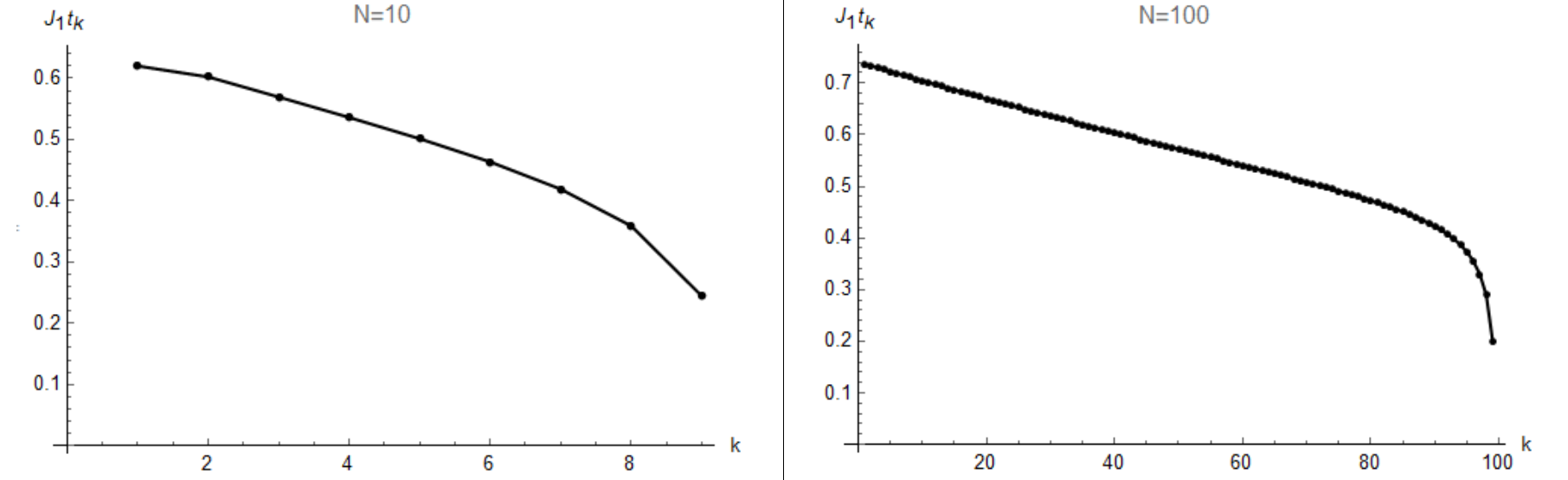}
 \caption{The time sequence of $Z_k$ pulses for generating W states on chains of length N=10 and N=100 sites.    $t_k$ is the time lapse between the $k-$th pulse and the $k+1$-th pulse. }
 \label{WSTimes}
 \end{figure}

 	The coupling strengths are found from (\ref{JEq}) and the times from (\ref{Tau_k+1}). The results are shown in figures (\ref{WSFig}) and (\ref{WSTimes}) for chains of length 10 and 100.
 	
 \subsection{Gaussian-states}
 
\begin{figure}[h]
	\centering
	\includegraphics[width=1\textwidth]{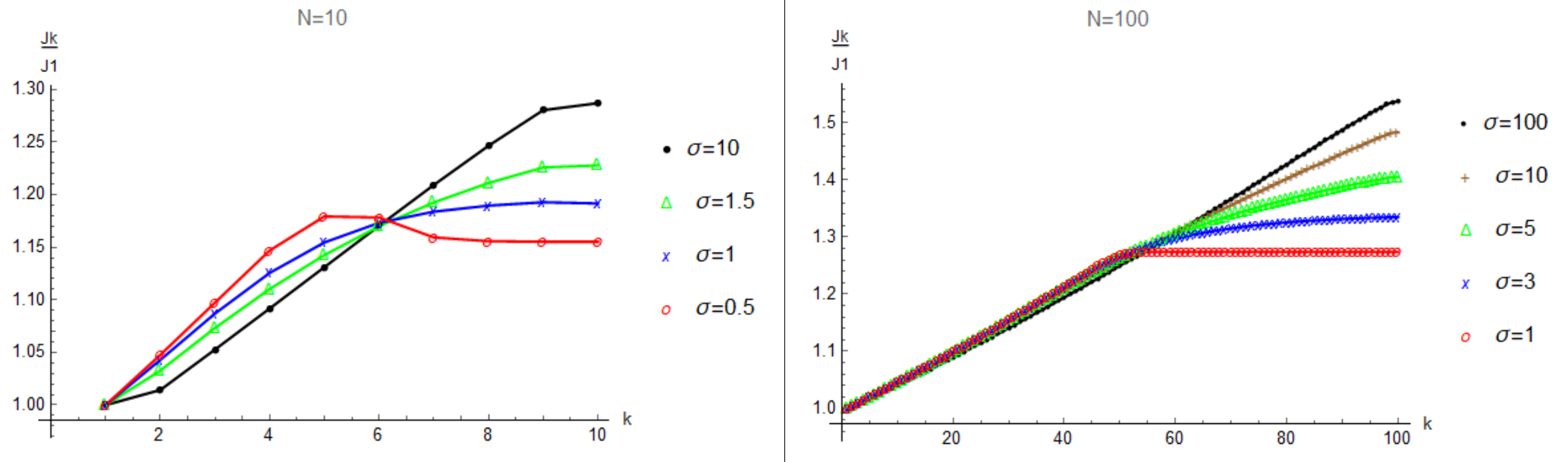}
	\caption{The coupling strengths for generating the Gaussian-states for n=10 and n=100.}
	\label{GSFig}
    \end{figure}

\begin{figure}[h]
   \centering
   \includegraphics[width=1\textwidth]{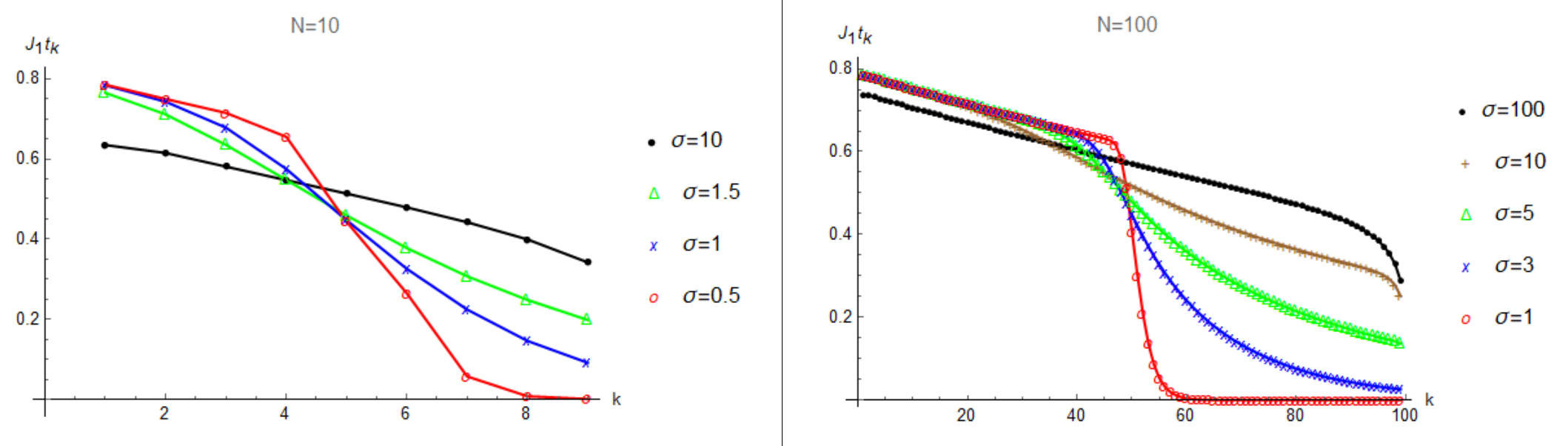}
   \caption{The time sequence of $Z_k$ pulses for generating Gaussian states with various widths.  $t_k$ is the time lapse between the $k-$th pulse and the $k+1$-th pulse. It is seen that after a while all the pulse can be applied simultaneously, specially for wave packets of small width. The reason is that after the localized packet has been generated, the excitation is effectively confined in the virtual chains in the middle of the chain. Hereafter, all the pulses on the empty chains on the right have no effect on the state.  }
   \label{GSTimes}
   \end{figure}

 	To generate a Gaussian state of a given width on the upper chain, we fix 
 	\be
 	\forall k\in\{1,2,...,N\}:\hmm P_{2k-1}=0\hmm,\hmm P_{2k}=\frac{e^{-\frac{(k-\frac{N+1}{2})^2}{2\sigma^2}}}{\sqrt{2\pi\sigma^2}}.
 	\ee
 	From which we find 
 	\be
 	 q_1=0\hmm,\hmm\forall k\in\{1,2,...,N-1\}:\hmm q_{2k}=q_{2k+1}=\frac{e^{-\frac{(k-\frac{N+1}{2})^2}{2\sigma^2}}}{2\sqrt{2\pi\sigma^2}}\hmm,\hmm q_{2N}=\frac{e^{-\frac{(\frac{N-1}{2})^2}{2\sigma^2}}}{\sqrt{2\pi\sigma^2}}.\nn
 	\ee
 	The coupling strengths and time sequences of pulses are shown in Fig.\ref{GSFig} and \ref{GSTimes} for N=10 and N=100 and for different values of $\sigma$.  
	
    By comparing Fig.\ref{WSFig} and Fig.\ref{GSFig}, we can see that the coupling strengths for generating W-states are very similar to the coupling strengths for generating Gaussian-states with a large standard deviation $\sigma$. It meets our expectation since Gaussian-states lead to W-states in the limit of a large standard deviation. By choosing the integers $m_k=0$ and $n_k=N$ in equations (\ref{JTauEq}) and (\ref{JtEq}), we have kept all the coupling constants finite and within the bounds provided in (\ref{BB1} and \ref{BB2}).
   
   \section{The sensitivity of the scheme against the timing of pulses}\label{error}
\begin{figure}[h]
   	\centering
   	\includegraphics[width=1\textwidth]{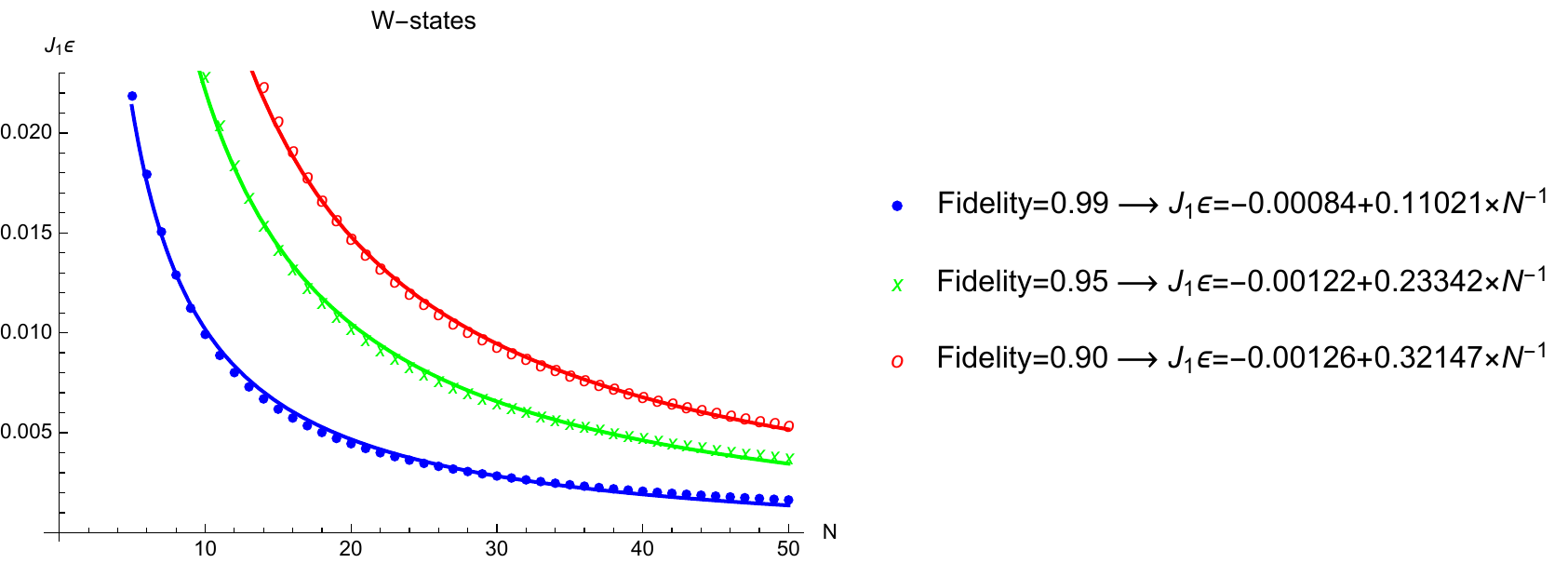}
   	\caption{(Color online)The fidelity of the generated W-state with the ideal W-state for chains of different lengths. The blue line separates the plane into regions of fidelity higher than 0.99 (below the curve) and lower than 0.99 (above the curve). This shows, for example that for chains of length $N=10$ and $20$ there is a tolerance of approximately $J_1\epsilon$ equal to 0.010 and  0.005 respectively. For lower fidelity, the green and red curves,  this tolerance naturally becomes higher. The curves from bottom to top are, blue, green and red. Note the nice scaling of the tolerance with the length of the chain written on the right-hand side of the plot.  Note that $J_1
   	\epsilon$ is the dimensionless quantity which should be tuned in order to attain a fidelity.  }
   	\label{wserr}
   \end{figure}

\begin{figure}[h]
	\centering
	\includegraphics[width=1\textwidth]{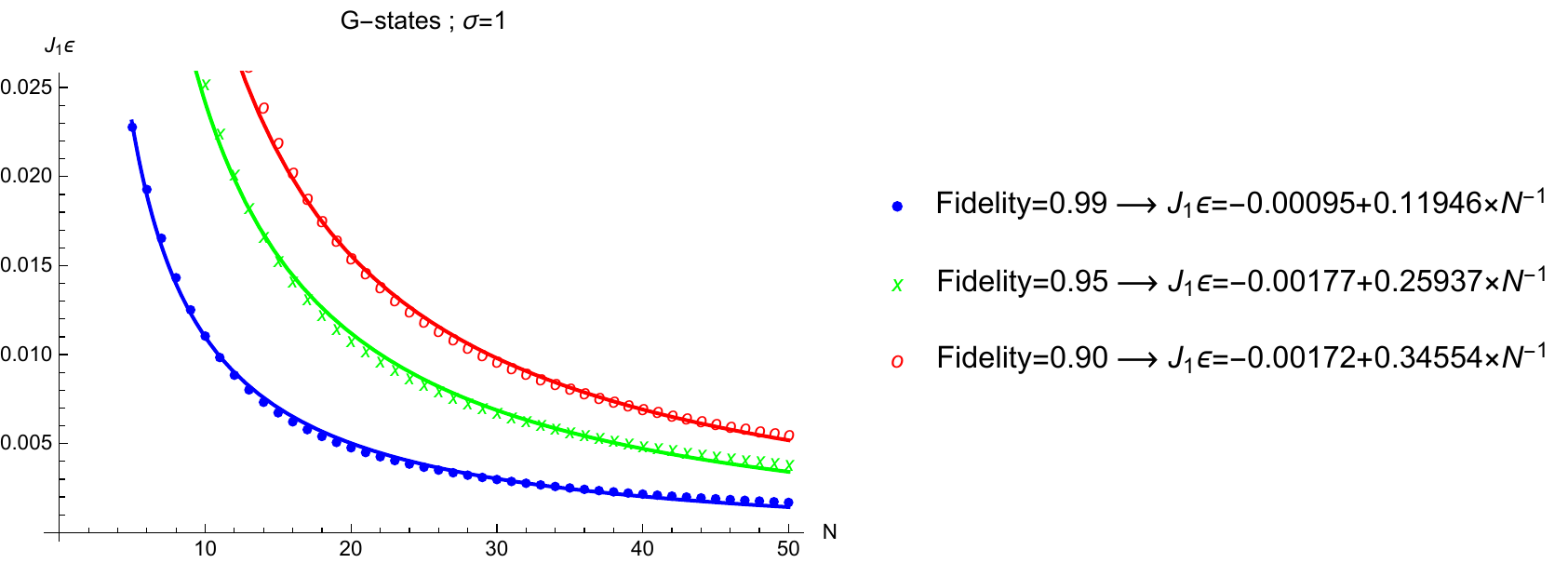}
	\caption{(Color online) Exactly the same description of the figure (\ref{wserr}) applies for this plot with the replacement of W-state with Gaussian states with $\sigma=1$. Only the curves for $\sigma=1$ are shown, the curves for other values of $\sigma$ are similar in shape with slightly different numerical factors.   }
	\label{gserr}
\end{figure}

   As equation (\ref{Tau_k+1}) shows, it seems that the exact states which are  produced depend  
   very much on the precise timing of the applied pulses. It is thus natural to ask how sensitive  this scheme is  with respect to this timing? What happens if the pulses are not applied exactly at the times demanded by Eq. (\ref{Tau_k+1}). We have done a detailed analytical treatment of this problem. However, reporting the details is not so illuminating and instead, we report the basic idea and the final numerical results. To simplify the analysis, let us assume that the times of free dynamics in all virtual 2-chains are dilated  or contracted by an amount $\epsilon$.    This means that there is an accumulative error in the time of all pulses, that is, the first pulse is applied with an offset error of $\epsilon$, the second pulse with an offset error of $2\epsilon$, the third pulse with an offset error of $3\epsilon$, and so on. Our intuitive reasoning that this  type of error, instead of a random error taken from a distribution, is the worst error that may happen  is the following.  The whole purpose of the pulses is to transfer an excitation from a virtual chain to the next virtual chain on the right time and if this transfer is delayed in each virtual chain, there comes a time where no excitation is in the middle of the chain to be transferred to the right end of the chain. In this case, the excitation will be trapped in some part of the left-hand side of the chain and goes back and forth in the virtual chains by the natural dynamics of these short chains. In this way, consecutive delays in these transfers hinders the desirable distribution of the excitation on the whole chain. 
   We have calculated the fidelity of the resulting state with the ideal state  generated by exactly applied pulses. The results are shown in figure (\ref{wserr}) for the W-state and in figure (\ref{gserr}) for the Gaussian state. The interesting point is the $\frac{1}{N}$ scaling of the required precision $\epsilon$  with the length of the chain for both types of states. 
   
	\section{Conclusion} \label{Conc}
	
	In this work and inspired by a technique first introduced in \cite{RossKay} and further developed in \cite{SAK}, we could exactly determine the coupling constants of a quasi-one-dimensional chain which is capable to generate any arbitrary singe excitation state. Instead of local magnetic fields $B_k$ which should be tuned along with the coupling constants $J_k$ in the work of \cite{Kay}, we had to use local pulses which have to be applied at definite times. By decomposing the chain into non-interacting virtual chains of length two whose dynamics is a simple rotation, we could exactly generate any single excitation state. Examples of $W-$ states and Gaussian states were studied the results of which are shown in figures \ref{WSFig} and \ref{GSFig}. Although the chain seems to be quasi-dimensional and  of a particular geometry, we can confine the state entirely on the upper chain and use the lower chain as an ancillary chain which is empty at the end of the process.

	\section{Acknowledgements} 
	We would like to thank members of the QIS group in Sharif or their valuable comments when this work was presented by one of the authors. This research was partially supported by a grant no. 96011347 from the Iran National Science Foundation. The work of V. K. was also partially supported by a grant from the research grant system of Sharif University of Technology.  
	
	{}


\begin{thebibliography}{}
		\bibitem{Bose1} S. Bose, Phys. Rev. Lett. \textbf{91} 207901 (2003).
		\bibitem{Bose2} S. Bose, Contemp. Phys. \textbf{48}, 13 (2007).
		\bibitem{ChDEL} M. Christandl, N. Datta, A. Ekert, and A. J. Landahl, Phys. Rev. Lett. \textbf{92}, 187902 (2004). M. Christandl, N. Datta, T. C. Dorlas,	A. Ekert, A. Kay, and A. J. Landahl, Phys. Rev. A \textbf{71}, 032312 (2005).
		\bibitem{BuB} D. Burgarth and S. Bose, New Journal of Physics 7, \textbf{135} (2005), ISSN 1367-2630.
		\bibitem{Kay1} A. Kay, Physical Rev. A \textbf{73}, 032306 (2006).
		\bibitem{FPK1} C. Di Franco, M. Paternostro, and M. S. Kim, Phys.
		Rev. Lett. \textbf{101}, 230502 (2008). 
		\bibitem{FPK2} C. Di Franco, M. Paternostro, and M. S. Kim, Phys. Rev. A \textbf{81}, 022319 (2010).
		\bibitem{BaB} A. Bayat and S. Bose, Phys. Rev. A \textbf{81}, 012304 (2010).
		
		\bibitem{Kay2} A. Kay, Int. J. Quantum Inform. \textbf{8}, 641 (2010).
		\bibitem{Jur} P. Jurcevic, et al, Nature 511, 202 (2014).
		 \bibitem{lat} J. I. Latorre, E. Rico, G. Vidal, Quant.Inf.Comput. 4 (2004) 48-92.
		 \bibitem{BACVV} L. Banchi, T. J. G. Apollaro, A. Cuccoli, R. Vaia, and P.Verrucchi, Phys. Rev. A \textbf{82}, 052321 (2010).

		\bibitem{Rus} R. Raussendorf, D. E. Browne, H. J. Briegel, Phys. Rev. A 68, 022312 (2003).
		\bibitem{BeB} S. C. Benjamin and S. Bose, Phys. Rev. Lett. \textbf{90}, 247901 (2003).
		\bibitem{KE} A. Kay, and M. Ericsson, New J. Phys. \textbf{7}, 143 (2005).
		\bibitem{RossKay} P. J. Pemberton-Ross and A. Kay, Phys. Rev. Lett. \textbf{106},	020503 (2011).
		\bibitem{SAK} V. Karimipour, M. Sarmadi Rad and M. Asoudeh,Phys. Rev. A \textbf{85}, 010302 (2012).
		\bibitem{AK} M Asoudeh, V Karimipour,  Quantum information processing 13 (3), 601-614.	
		\bibitem{Kay} A. Kay, New. J. Phys. \textbf{19}, 043019 (2017).
		\bibitem{jos} M. Paternostro, G. M. Palma, M. S. Kim, G. Falci, Phys. Rev. A 71, 042311 (2005), A. Romito, R. Fazio and C. Brudo, Phys. Rev. B 71, 100501(R) (2005), A. Lyakhov and C. Bruder, New J. Phys. 7, 181 (2005).

	\end{thebibliography}
\end{document}